\documentclass{article}

\usepackage{PRIMEarxiv}

\usepackage[utf8]{inputenc} 
\usepackage[T1]{fontenc}    
\usepackage{hyperref}       
\usepackage{url}            
\usepackage{booktabs}       
\usepackage{amsfonts}       
\usepackage{amssymb}
\usepackage{textcomp}
\usepackage{nicefrac}       
\usepackage{microtype}      
\usepackage{lipsum}
\usepackage{nomencl}
\usepackage{framed}
\usepackage{multicol}
\usepackage{fancyhdr}       
\usepackage{graphicx}       
\graphicspath{{media/}}     
\usepackage{siunitx}
\usepackage{gensymb}
\usepackage[sort&compress,numbers]{natbib}
\makenomenclature
\title{Heat transport during drop impact onto a heated wall covered with an electrospun nanofiber mat: The influence of wall superheat, impact velocity, and mat thickness
}

\author{
  A. Gholijani, T. Gambaryan-Roisman, P. Stephan \\
  Institute of Technical Thermodynamics, Technische Universität Darmstadt, \\
  Peter-Grünberg-Straße 10, 64287 Darmstadt, Germany \\
  \texttt{gtatiana@ttd.tu-darmstadt.de} \\
}

\begin{document}
\maketitle

\begin{abstract}
Nanofiber surface coating is a promising method for the enhancement of heat transfer during spray cooling. In the present work, the drop dynamics as well as local and overall heat transfer during single drop impact onto a heated wall covered with a nanofiber mat are investigated to obtain insight into the mechanisms governing the heat transport enhancement. The influence of wall superheat, drop impact velocity, and mat thickness on the hydrodynamics and heat transfer from the heated wall to the fluid is studied. Polyacrylonitrile (PAN) was electrospun on a heater surface to manufacture the nanofiber mat coatings with thicknesses between $18$ and 80~{\textmu}m; the cross-sectional diameter of the fibers ranged from $200$ to $300$ nm. The experiments were conducted inside a temperature-controlled test cell with a pure vapor atmosphere maintained with refrigerant FC$-72$ (perfluorohexane). The temperature field at the solid-fluid interface was observed with a high-speed infrared camera, and the heat flux field was derived by solving a three-dimensional transient heat conduction equation within the substrate. The dynamic evolution of the contact radius was derived using image analysis. The presence of the nanofiber mat on the heater surface suppresses the drop receding phase due to the pinning of the contact line at the end of the spreading phase. At the initial impact stage, a high heat flux is transferred to the drop in the case of a bare heater, while the nanofiber mat on the heater surface lowers the heat flux. At the later stage of impact, two different scenarios are observed depending on the wall superheat and drop impact velocity: scenario (I), in which the liquid drop penetrated the porous nanofiber mat and made contact with the heater surface; and scenario (II) in which the vapor produced inside the pores of the nanofiber mat prevented the liquid drop from touching the heater surface. At a certain point in time after impact, the energy transferred from the nanofiber-coated surface to the liquid exceeds that of the uncovered heater owing to the larger drop footprint. If scenario (I) occurs, then the total transported heat increases significantly compared with the drop impact on a bare substrate.
\end{abstract}


\begin{table*}[!t]
    \centering
    \textbf{\large{Nomenclature}}\\\vspace{0.5cm}
    \begin{tabular}{|m{1.5em} m{19em} m{1.5em} m{19em}|} 
    \hline
    &&&\\
    	  {$c$}&{specific heat of liquid (J kg$^{-1}$ K$^{-1}$)} & {$Re$}&{Reynolds number}\\
      {$D_0$}&{drop diameter (m)}&{$S$}&{spreading ratio}\\
     {$E$}&{cumulative heat (J)}&{$T$}&{temperature (K)}\\
	{$E^*$}&{dimensionless cumulative heat}&{$t$}&{time (s)}\\
	{$h$}&{mat thickness (m)}&{$u_0$}&{impact velocity (m s$^{-1}$)}\\
	{$h_{lv}$}&{enthalpy of vaporization (J kg$^{-1}$) }&{$We$}&{Weber number}\\
    {$Ja$}&{Jakob number}&{$\Delta T$}&{wall superheat, K}\\
{$\dot{Q}$}&{heat flow (W)}&{$\mu$}&{dynamic viscosity of liquid (kg m$^{-1}$ s$^{-1}$)}\\
{$Q^*$}&{dimensionless heat flow}&{$\nu$}&{kinematic viscosity of liquid (m$^2$ s$^{-1}$)}\\
{$\dot{q}$}&{heat flux (W m$^{-2}$)} & {$\rho$}&{density of liquid (kg m$^{-3}$)}\\
{$R$}&{drop spreading radius (m)}       & {$\sigma$}&{surface tension (N m$^{-1}$)}\\
&&  {$\tau$}&{dimensionless time}\\
&&&\\
\hline
\end{tabular}
\end{table*}

\section{Introduction}
Miniaturization and breakthroughs in electronic, optical, and radiological components require high-performance cooling methods, which have been comprehensively reviewed in \cite{Yarin2009, zhang2021review}. To date, various cooling methods such as air cooling, piezo fans, liquid jet cooling, heat pipes, and spray cooling have been applied to enhance the heat transfer from such devices. Spray cooling is considered one of the most effective heat removal methods with a high probability of reaching its maximum efficiency.

One approach to significantly enhance spray cooling efficiency is surface modification. A low-cost method of surface modification is coating the surface with an electrospun nanofiber mat. This method has been shown to drastically increase the cooling rate during the drop impact cooling \cite{Srikar,Weickgenannt1,Weickgenannt2,Sinha-Ray2011,Sinha-Ray20141,Sinha-Ray20142, park2023drop}, pool boiling \cite{Jun,Sahu,Sinha-Ray2017}, and flow boiling in mini-channels \cite{Freystein}. Electrospun nanofiber mats consist of nonwoven polymer nanofibers that are randomly oriented inside the porous layer. The mats are highly porous and permeable; the size of the inter-fiber pores is in the order of several micrometers, and the diameter of the fibers is approximately several hundred nanometers \cite{Yarin2007,Greiner,Reneker,Agarwal}.

Drop impact on an impermeable surface is associated with various outcomes, such as drop deposition, splashing, or bouncing, depending on the wall temperature, impact parameters (e.g., impact velocity and impact diameter), surface morphology, and wettability \cite{Breitenbach20181}. The outcomes of water drop impact on surfaces covered by nanofiber mats with a thickness of about  $200~${\textmu}m have been studied experimentally at isothermal conditions in \cite{Lembach}. It has been found that the threshold for splashing of a water drop on a nanomat-coated substrate was shifted towards higher impact velocities in comparison to flat, smooth impermeable surfaces. In addition, the drop behavior in the deposition regime significantly changes for surfaces with porous nanofiber coatings in comparison with impermeable surfaces. In case of drop deposition on impermeable surfaces, the drop impact is followed by drop spreading, receding, and sessile drop phases. The drop receding phase was shown to be suppressed during the water drop impact on nanofiber mats, as the contact line was pinned at the end of the spreading phase. In addition, it was observed that the liquid gradually impregnated the pores of the mat outside the area covered by the drop \cite{Lembach}. The supression of the drop receding phase and the impregnation of the pores (imbibition od the liquid into the porous layer) are advantageous for heat transfer applications, since they lead to increasing of the contact area between the substrate and the liquid, which promotoes the sensible heat transfer between the solid and the liquid, and the increasing of the liquid-gas interface area, which promotes evaporation.   

If a liquid drop impacts a heated wall, the wall temperature affects both the impact dynamics and the heat transfer performance. \citet{Breitenbach20181} distinguished following five heat transfer regimes (in the order of increasing wall temperature): (i) evaporation below the onset of boiling, (ii) nucleate boiling, (iii) transition boiling, (iv) thermal atomization, and (v) film boiling. The emergence of one of these regimes depends, in addition to wall temperature, on the impact Reynolds and Weber numbers, and the substrate properties.

\citet{Weickgenannt1,Weickgenannt2} studied the impact of drops onto a 50~{\textmu}m-thick heated stainless steel foils coated with electrospun polyacrylonitrile (PAN) nanofiber mat with a thickness in the range from 0.15~mm to 1.5~mm. The measurements were conducted in ambient air. The temperature distribution at the backside of the foil was measured using an infrared (IR) camera. In \cite{Weickgenannt1},  water was used as a test fluid, and the initial foil temperature was varied between 60°C and 260°C.  In \cite{Weickgenannt2}, water and ethanol drops have been used, and the maximal initial foil tempeature reached 300°C. In all experiments, heat transfer enhancement has been observed for the coated foils im comparison with the bare steel foils. This heat transfer enhancement manufested itself in lower foil temperatures after the drop impact and in shorter drying times. At moderate initial foil temperatues, the suppression of the drop receding phase and the imbibition of the liquid into the porous coating beyond the maximal spreading radius were responsible for the intensification of heat transfer. At high wall temperatures, using of coated substrates lead to change of the themal regime. At the initial foil temperatures of 220°C and 300°C, the impact of ethanol drops resulted in a film boiling regime. This effect is known as the Leidenfrost phenomenon, which manifests itself in a development of a vapor film between the wall and liquid drop, hindering high heat transfer \cite{Leidenfrost}.  \citet{Weickgenannt2} discovered that nanofiber coating completely suppressed the Leidenfrost effect. The \textquotedblleft \space inverse-Leidenfrost \textquotedblright \space effect caused by nanofiber coating increases the rate of heat removal from the surface to the liquid drops tremendously. 

The results reported in \cite{Srikar, Lembach, Weickgenannt1, Weickgenannt2} have been collected for relatively thick nanomat coatings. \citet{Heinz2021} performed experimental study of spreading, imbibition and evaporation of ethanol drops on silicon surfaces coated by electrospun PAN nanofiber mats with thicknesses in the range from 4~{\textmu}m to 42~{\textmu}m, which required a shorter manufacturing time and lower amount of the polymer and the solvent. It has been shown that that the maximal area of imbibed region increased with increasing of the coating thicknesss from 4~{\textmu}m to 14~{\textmu}m, and the further increasing of the coating thickness did not affect the imbibition area and evaporation time. Although the experiments have been performed without heating the substrates, the results presented in \cite{Heinz2021} allowed to suggest that applying the nanofiber mat in the thickness range 15 - 25~{\textmu}m can efficiently promote the heat transport.

The reported quantitative heat transfer data related to drop impact on surfaces with nanomat coatings have been collected at a time scale of several seconds, which is a characteristic time of imbibition of millimeter-size drops of water or ethanol. The heat transfer during the initial spreading of the drop on the surface has not been quantified. However, it is known that the highest heat transfer rates are achieved during the inertia-driven drop spreading after the impact \cite{lee2001time, herbert2013local, herbert2013influence}. The porous coating constitutes a hydrodynamic resistance and can therefore defer the contact between the cool liquid and a heated surface. The effect of this resistance on heat transport during the initial drop spreading has not been yet sufficiently studied. In addition, the experiments reported so far have been performed in an atmosphere of ambient air, whereas evaporation of liquid within a saturated vapor atmosphere is more efficient and is therefore more promising for cooling applications.    


Recently, the results of experimental studies on hydrodynamics and heat transport during impact of one or several drops of the refrigerant FC-72 on a hot surface in a pure vapor atmosphere have been reported \cite{Gholijani1, Gholijani2, gholijani2022experimental}. In these works, the design of the heater in combination with the infrared thermography allowed a simultaneous determination of the temperature and heat flux fields at the solid-fluid interface with a high temporal resolution. In the present study, we apply the experimental methodology used in \cite{Gholijani1, Gholijani2, gholijani2022experimental} to investigate the drop dynamics and local heat transport during drop impingement on hot surfaces covered with nanofiber mats, with pure vapor as the surrounding atmosphere. The experiments are conducted under drop deposition regime (small impact Reynolds and Weber numbers) and evaporation regime below the onset of boiling (moderate wall superheats). The influence of wall superheat and drop impact velocity in the presence of the nanofiber mat, and the influence of mat thickness on the drop dynamics as well as the local and overall heat transfer from the heated wall to the liquid phase are studied.

\section{Experimental method}
\subsection{Experimental setup}

\begin{figure}
\centering\includegraphics[width=1\linewidth]{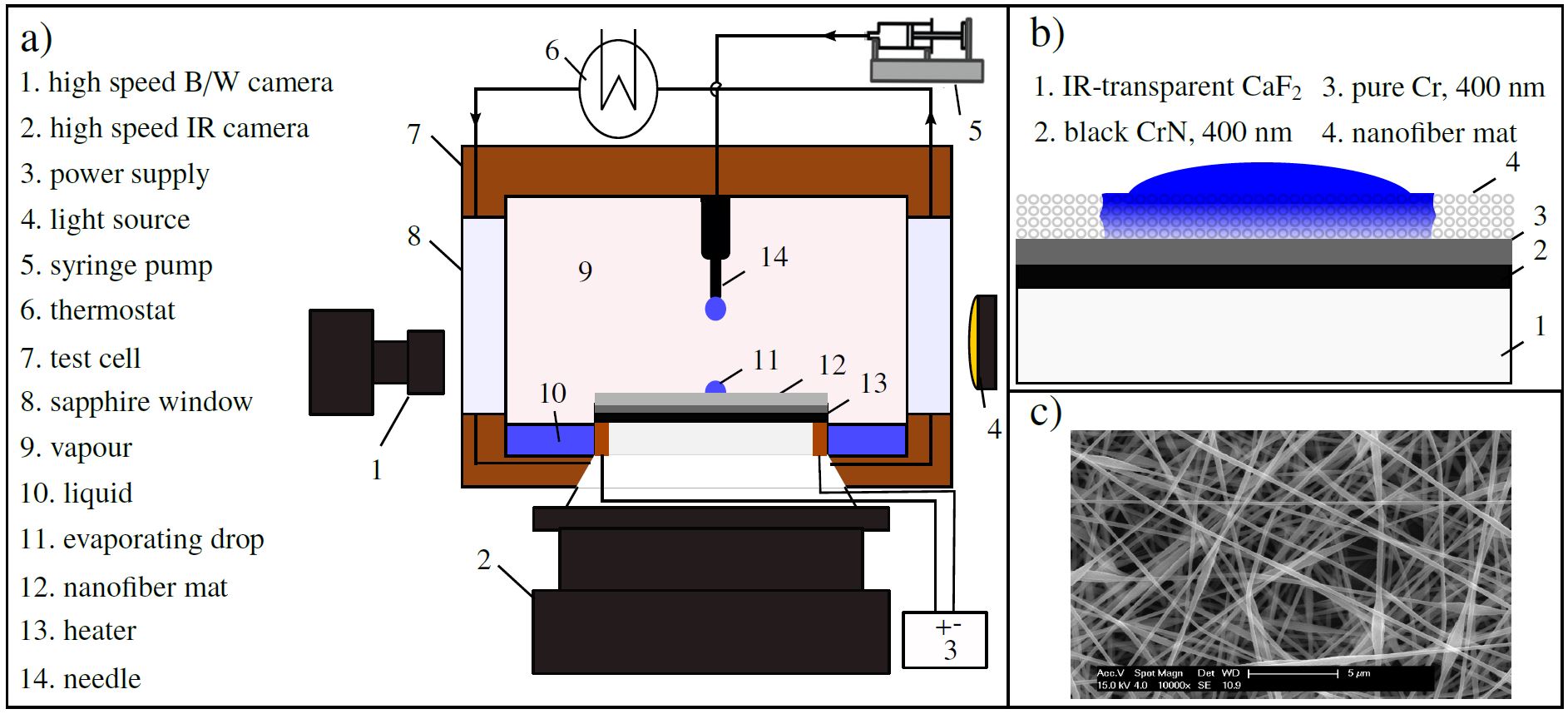}
\caption{a) schematic of the setup, b) schematic of the heater, and c) SEM image of the nanofiber mat}
\label{Setup}
\end{figure}

Figure \ref{Setup}a shows the schematic of the drop impact experimental setup. The core of the setup is a sealed temperature-controlled test cell containing a fluid maintained under saturation conditions. The refrigerant FC$-72$ (perfluorohexane) with a saturation temperature of $56.6^{\circ}$C at atmospheric pressure was used as the working fluid. The pressure inside the cell is determined by the fluid temperature and can be varied and controlled by pumping liquid with a controlled temperature through the channels drilled into the wall of the cell. Prior to filling the test cell with the working fluid, the non-condensable gases were extracted from the liquid using a degassing experimental setup described in \cite{Fischer}. Afterward, the cell was vacuumed from the air and filled with the degassed working fluid to ensure pure saturation conditions. 

A syringe pump was employed to generate a single drop by pumping the degassed fluid from the reservoir to the cell. The drop diameter is highly reproducible, as the detachment of the drop is governed by the balance between gravity and surface tension forces, both of which remained constant during the experiments. In this study, a Gauge $30$ needle was used, which resulted in a constant drop diameter of $0.95$~mm. The impact velocity have been varied by installing the needle at various heights over the substrate. The impact velocities ranged from $0.32$ to $0.58$~m~s$^{-1}$. The input heat flux to the heater was controlled by passing an electric current through the heater. The measured input heat flux ranged from $850$ to $5200$ W m$^{-2}$, which corresponds to the wall superheat (the temperature difference between the wall and the saturation temperature) between $1.2$ and $17$ K. 
 
A schematic of the heater is shown in Fig.~\ref{Setup}b. The heater consists of a $4$~mm thick CaF$_2$ substrate, which is transparent to an infrared (IR) radiation. Two thin layers were deposited onto a CaF$_2$ substrate through the physical vapor deposition (PVD) process \cite{Slomski}. A $400$ nm CrN layer was applied to increase the surface emissivity, and a $400$ nm pure Cr layer was applied  as a resistance heater. Because the thicknesses of both Cr and CrN layers are small, the energy storage and thermal resistance of these layers are negligible. The arithmetic mean deviation, maximum valley depth, maximum peak height, and maximum height of the substrate surface profile measured via confocal microscopy are $5$, $15$, $15$, and $30$ nm, respectively. Therefore, the bare heater surface was smooth. Under isothermal conditions, the measured static contact angle of FC-$72$ on the surface is less than 3\degree. 

To create a porous coating, the heater substrate was covered with a nanofiber mat consisting of randomly oriented nanofibers generated through electrospinning of $5$ wt$\%$ polyacrylonitrile solution (M$_w$ = 150 kDa) in N,N-dimethylformamide (DMF). Figure \ref{Setup}c shows the SEM image of the generated nanofibers. The nanofibers were electrospun onto the heater substrate for a particular time span to achieve distinct mat thicknesses. The mat thicknesses, measured using a confocal microscope, are $18$, $22$, $51$, and $80$~{\textmu}m. The fiber diameters ranged from $200$ to $300$ nm. 
   
Optical access to the drop was obtained through transparent windows embedded on the sides of the cell. A high-speed black and white (B/W) camera was installed on the side of the cell to record the drop impact parameters. Its spatial resolution and frame rate were set to $10.87$~{\textmu}m/pixel and $1000$ Hz, respectively. The drop size and speed at the moment of impact were evaluated through post-processing of the captured B/W images. A high spatial and temporal resolution IR camera was installed at the bottom of the cell to capture the temperature field on the backside of the CrN layer. The spatial and temporal resolutions of the IR camera are $29.27$~{\textmu}m/pixel and $1$ ms, respectively. Both cameras were temporally synchronized. After each experimental run, an in situ calibration of the IR signal versus the temperature for each pixel was performed. This was achieved by pressing a copper block with a known temperature on the heater surface and capturing IR images at various temperature intervals.

\subsection{Data reduction}
The main focus of this work is the evaluation of the local heat flux from the heater to the fluid domain, followed by the assessment of the heat flow and drop spreading radius. The local heat flux was derived from the temperature field by solving the three-dimensional transient heat conduction equation using the finite volume method in the computational fluid dynamics toolbox OpenFOAM. The details of this procedure and the accuracy of this method are given in \cite{Fischer20121,Fischer20122,Gholijani1}.

The temporal evolution of the drop spreading radius was derived by post-processing the evaluated heat flux images. 
The method employed to determine the drop spreading radius is described in detail in \cite{Gholijani1}. The corresponding heat flow was evaluated by integrating the local heat flux over the drop spreading radius.

The drop diameter prior to impact and the impact velocity were derived via post-processing of the side-view images captured with the high-speed B/W camera. The drop diameter was evaluated using a three-dimensional volume integration method. This method is based on the fact that the two-dimensional images of the drop captured by the high-speed camera using a telecentric lens are orthogonal projections of the drop. In this method, the drop was assumed to be symmetric about the vertical axis. The corresponding three-dimensional representation of each pixel inside the image is a ring with a square cross-section. The sum of the evaluated volumes of the ring was obtained to compute the volume of the drop. Finally, the equivalent drop diameter was derived from the volume of the drop.
	
The velocity of the drop for five frames before impact was used to determine the impact velocity. The velocity has been determined as the ratio between the displacement of the drop center and the time interval, which was $1$ ms. In addition, the distance between the center of the drop in the frame immediately preceding the impact and the heater surface was assessed. Then, a square root function based on the equation of mechanical energy conservation was fitted to the calculated velocities. Finally, the impact velocity was evaluated via curve extrapolation. In this study, the drop diameter remained constant at $0.95$~mm, while the impact velocities ranged from $0.45$ to $0.58$~m~s$^{-1}$. 

The noise equivalent temperature difference (NETD) and noise equivalent heat flux difference (NEHFD) methods were used to evaluate the measurement uncertainties of the heat flow \cite{Fischer,Gholijani1,Gholijani2}. NETD is a function of the temperature and optical setup. NEHFD represents the heat flux uncertainty and can be evaluated from the NETD. The calculated heat flux uncertainty with this method is approximately $7500$ W m$^{-2}$. The measurement uncertainty of the contact line radius corresponds to the size of two pixels taken from image processing.

The uncertainty in the determination of the heat flow is governed by the uncertainties of the contact line radius and heat flux. The maximum heat flow uncertainty for the impact of each drop occurred at the instant of maximal spreading. For example, the maximum heat flow uncertainty during the impact onto a heater without a nanofiber mat at $u_0 = 0.58$~m~s$^{-1}$ and $\Delta T = 9.6$ K is $0.13$ W (see Fig.~\ref{BN-HF}, top). The measurement uncertainty of the liquid temperature inside the cell is $0.6$ K.

\section{Results and discussion}
In this section, the drop dynamics and heat transport on heated substrates coated by nanomats are analysed. First, the behavior of the drops on the coated heater is compared with the behavior of a drop on a bare heater. After that, the influence of wall superheat and drop impact velocity, as well as the influence of mat thickness on the drop dynamics and heat transport are discussed.
 
\subsection{Influence of nanofiber mat on drop dynamics and heat transport}\label{Nan}
\begin{figure}
\centering\includegraphics[width=1\linewidth]{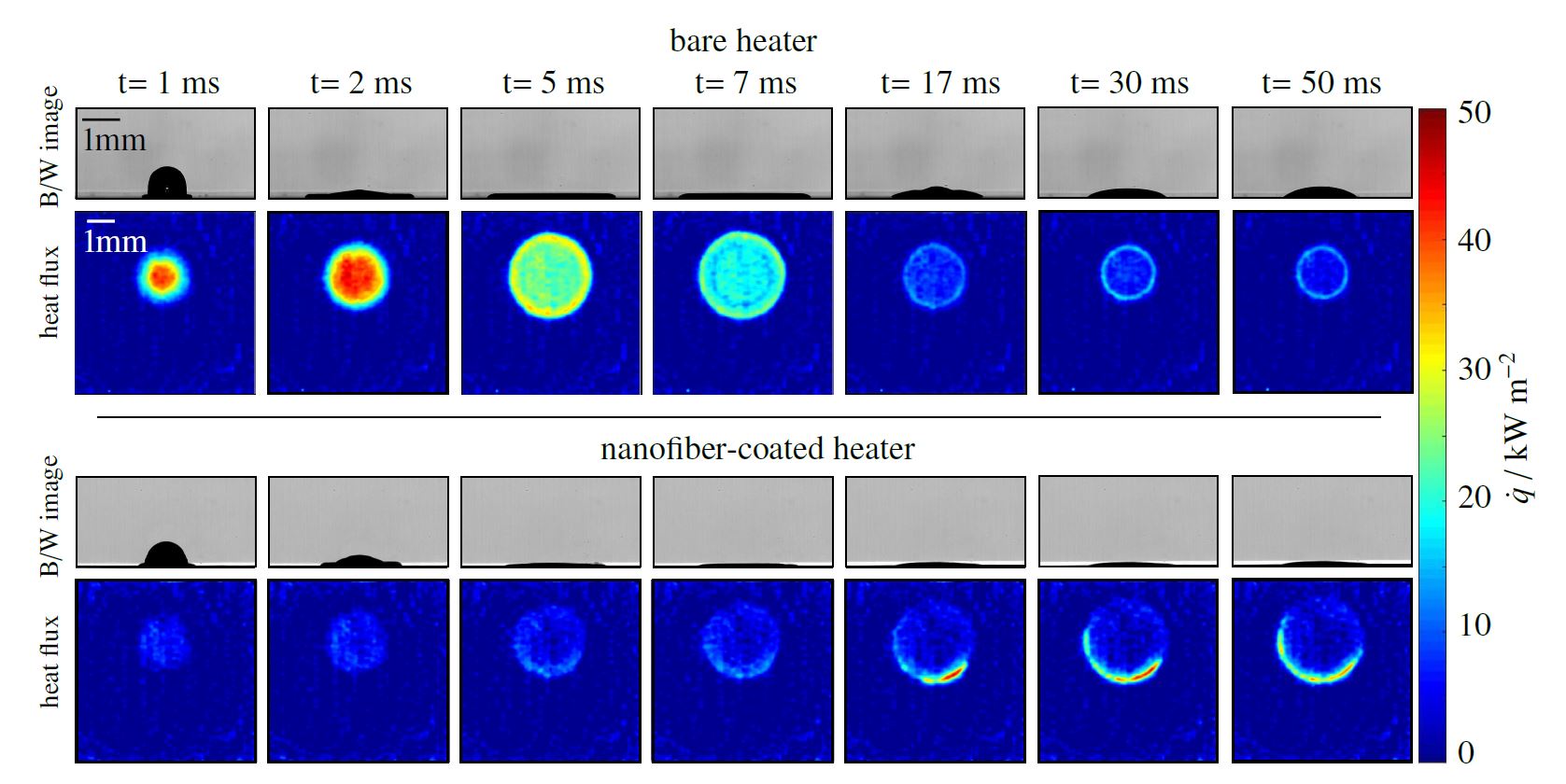}
\caption{B/W images showing the side-view of the drop impact and spreading, and the corresponding heat flux profiles at the wall surface within the first $50$ ms after impact ($D_0= 0.95$~mm, $u_0= 0.45$~m~s$^{-1}$, and $\Delta T= 9.1$ K), top: without nanofiber mat, bottom: with nanofiber mat.}
\label{Comp1}
\end{figure}

The typical temporal evolution of the heat flux distribution evaluated from the IR images and the side-view of the drop shape evolution captured by the B/W camera in the presence and absence of the nanofiber mat are presented in Fig.~\ref{Comp1}. The thickness of the nanofiber mat is $h= 22$~{\textmu}m. It can be seen that the impingement on the uncovered (bare) heater surface is divided into three phases: (i) the drop spreading phase, where the drop contact radius increases (for the measurement depicted in Fig.~\ref{Comp1}, the time span is $7$ ms); (ii) the drop receding phase, where the drop contact radius decreases (for the measurement depicted in Fig.~\ref{Comp1}, the time span is $20$ ms); and (iii) the sessile drop evaporation phase, where the liquid drop reaches the stationary stage. During the spreading phase, the hydrodynamics is governed by inertia, whereas during the receding phase, the hydrodynamics is governed by surface tension.

\begin{figure}
\centering\includegraphics[width=1\linewidth]{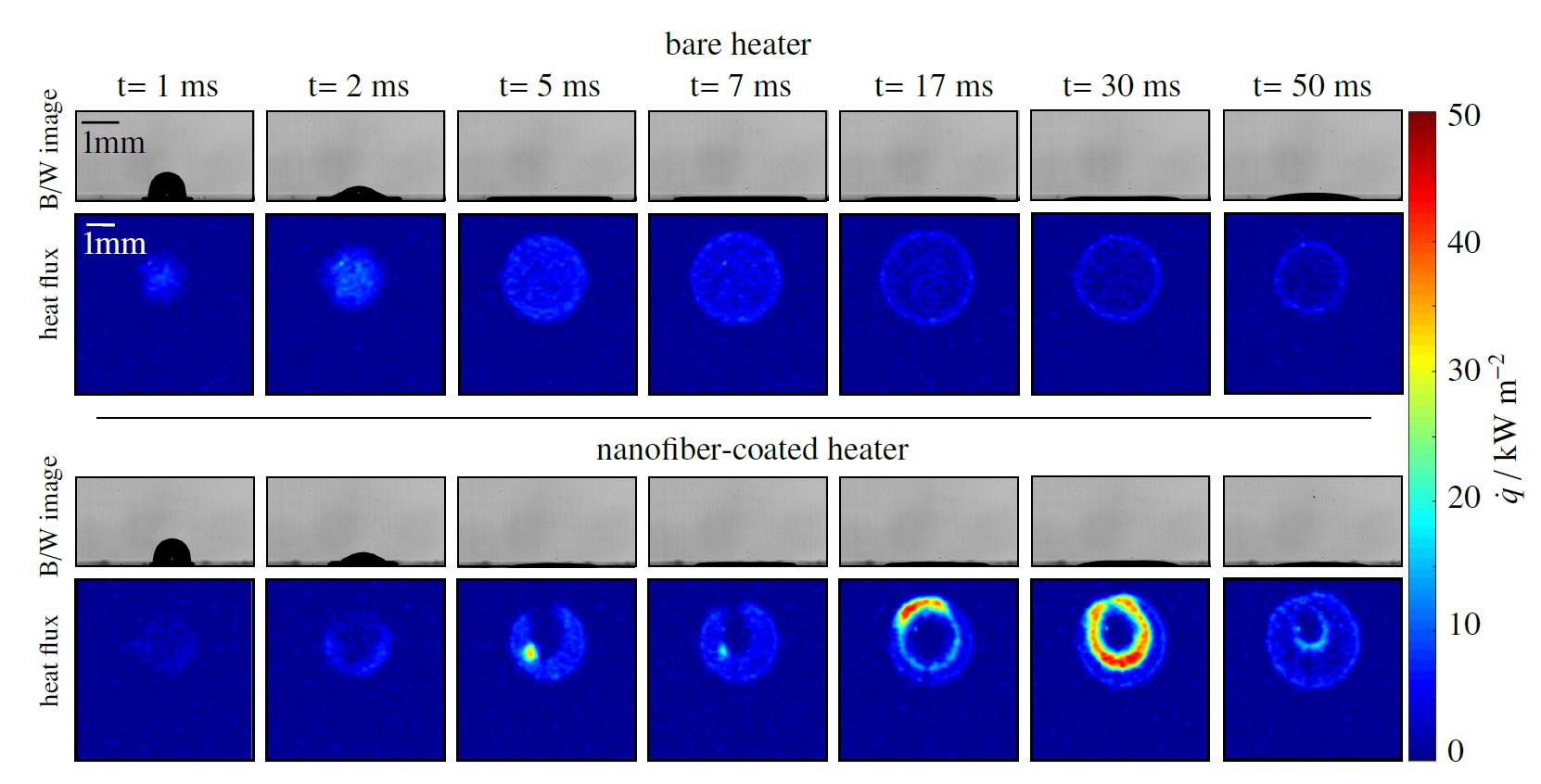}
\caption{B/W images showing the side-view of the drop impact and spreading, and the corresponding heat flux profiles at the wall surface within the first $50$ ms after impact ($D_0= 0.95$~mm, $u_0= 0.45$~m~s$^{-1}$, and $\Delta T= 2.2$ K), top: without nanofiber mat, bottom: with nanofiber mat.}
\label{Comp2}
\end{figure}


Similar to the bare surfaces, the inertial force leads to drop spreading over the nanofiber mat at the initial stage of impact. However, the spreading velocity is slower than that on the uncovered heater. Thereafter, the contact line is pinned, which leads to the suppression of the drop receding phase. Finally, the liquid drop slightly shrinks, and its spreading radius moderately decreases because of the liquid evaporation.

The drop impingement on an uncovered heater is accompanied by a high local heat flux immediately after the impact and during the initial spreading. The high heat flux is caused by the direct contact of the cold liquid with the hot heater surface. On the other hand, the presence of the nanofiber mat prevents the direct contact between the liquid and the heater surface at the early stage of impact. Two mechanisms can be responsible for the deferring of the contact. First, the hydrodynamic resistance created by the mat decelerates the liquid flow towards the heated substrates. Second, it is suggested that the cold liquid initially touches the top surface of the hot nanofiber mat and evaporates partially. The generated vapor is trapped inside the pores between the Cr layer and bulk liquid. This phenomenon, which is attributed to the \textquotedblleft skeletal\textquotedblright \space Leidenfrost effect, significantly weakens the heat flux because of the low thermal conductivity of the vapor \cite{Srikar}. In such a case, the drop first cools down the nanofibers, and heat is removed from the heater mainly through the nanofiber skeleton ($k_{FC-72,v} = 0.008$ and $k_{fiber}= 0.02-0.05$ W m$^{-1}$ K$^{-1}$ \cite{Sabetzadeh}).

After several milliseconds, two scenarios may occur depending on the wall superheat and impact velocity. Scenario (I): the liquid drop completely penetrates into nanofiber pores and reaches the solid heater surface (see Fig.~\ref{Comp2}). This scenario is realized with low wall superheat or high impact velocity. High heat flux areas appear at the solid-liquid interface at the drop footprint. The location of the high heat flux areas changes dynamically with time. It can be suggested that small areas at the drop footprint are periodically dried out and rewetted. Scenario (II) is realized if the wall superheat is sufficiently high and the impact velocity is low. A significant amount of vapor is generated and trapped inside the pores, which hinders the liquid drop from reaching the heater surface (see Fig.~\ref{Comp1}). As a result, the heat flux in this scenario is lower than that in scenario (I). 

\begin{figure}
\centering\includegraphics[width=1\linewidth]{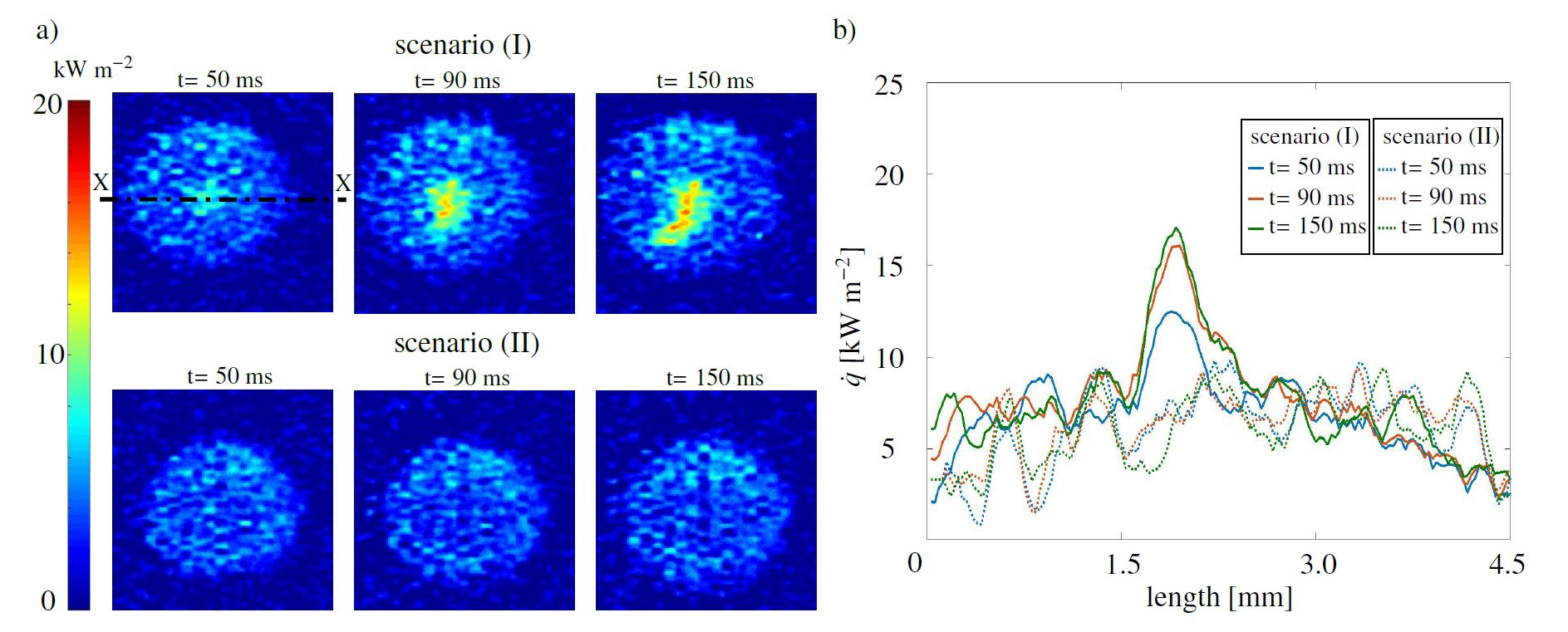}
\caption{a) Heat flux fields at $t = 50, 90$, and $150$ ms after impact for scenarios of (I) and (II); and b) heat flux distribution along the centerline X-X.}
\label{CompSc}
\end{figure}

In Fig.~\ref{CompSc}(a), the heat flux fields, and in Fig.~\ref{CompSc}(b), corresponding heat flux distributions along a line passing through the center of the drop footprint at $t = 50, 90$, and $150$ ms are displayed for scenarios (I) and (II). The scenarios (I) and (II) are realized with the impact velocities $u_0= 0.58$ and $0.45$~m~s$^{-1}$, respectively. The comparison was made during the sessile drop evaporation phase when the convection within the drop is negligible. The measured heat fluxes in scenario (I) show distinct regions of high heat flux which are not observed in scenario (II). In this example, the high heat flux region in scenario (I) occurred in the middle of the impacted area; however, the position of the high heat flux can be different in other measurements and varies over time.  

\begin{table}
 \caption{Dimensional and dimensionless impact parameters and wall superheat as well as mat thickness for scenarios (I) and (II).}
  \centering
  \begin{tabular}{lll}
    \toprule
   Parameters& Scenario (I) & Scenario (II) \\
    \midrule
		$D_0$ [mm]&0.95&0.95\\ 
		$u_0$ [m s$^{-1}$]&0.58&0.45\\ 
		$\Delta T$ [K]&9.6&7.0\\ 
		$h$ [{\textmu}m]&22&22\\ 
		Re [-]&1920&1550\\ 
		We [-]&60&39\\ 
		Ja [-]&0.12&0.09\\ 
    \bottomrule
  \end{tabular}
  \label{Sen}
\end{table}

\begin{figure}
\centering\includegraphics[width=0.6\linewidth]{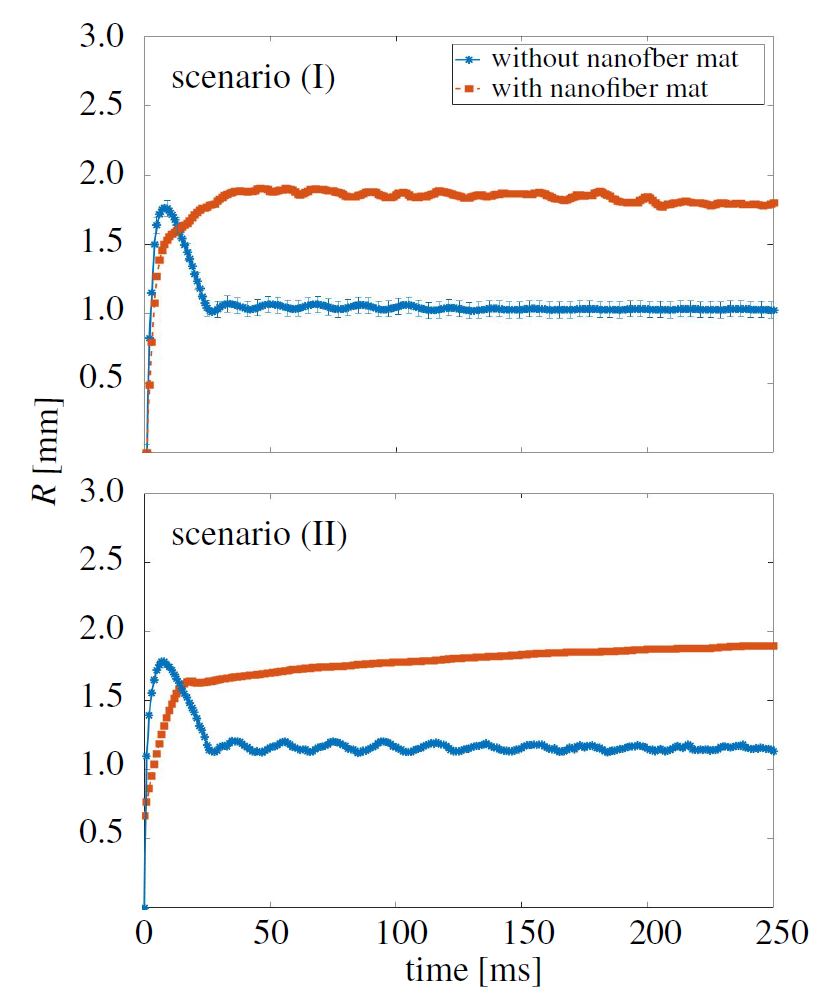}
\caption{Temporal evolution of the drop spreading radius on a heater without and with the nanofiber mat for scenarios (I) and (II) for the parameters listed in Table \ref{Sen}.}
\label{BN-Rad}
\end{figure}

Figure \ref{BN-Rad} illustrates the temporal evolution of the spreading radius in scenarios (I) and (II) for a heater covered with the nanofiber mat. The corresponding parameters are listed in Table \ref{Sen}. The impact Reynolds number (Re) expresses the ratio of the inertial and viscous forces:
\begin{equation}\label{eq:Re}
Re= \frac {\rho D_0u_0} {\mu},
\end{equation}
the Weber number (We) represents the ratio between the inertial and surface tension forces:
\begin{equation}\label{eq:We}
We= \frac {\rho D_0u_0^2} {\sigma},
\end{equation}
and the Jakob number (Ja) represents the ratio of the sensible heat to the latent heat transferred during the phase change process:
\begin{equation}\label{eq:Ja}
Ja= \frac {c\Delta T} {h_{lv}}.
\end{equation}

For comparison, the data for drop impact on an uncovered heater surface at the same impact parameters are shown in Fig.~\ref{BN-Rad}. In the case of an uncovered heater, the drop spreading phase lasts a few milliseconds with a relatively large spreading velocity. Then, drop receding occurrs until $t= 30$ ms. Finally, the drop oscillated several times until it reached its equilibrium stage.


{If the heater is covered with a nanofiber mat, then the drop spreading phase can be further divided into two phases regardless of whether scenario (I) or (II) occurs: drop spreading (i) with high spreading velocity at the initial stage of impact, and (ii) with low spreading velocity thereafter. The inertial force is the governing force for the high spreading velocity phase at the initial stage of impact. In scenario (I), the spreading velocity at the early stage of impact is close to the spreading velocity on the uncovered heater. Afterward (second spreading phase), the spreading slows down compared with the uncovered heater substrate. In scenario (II), the spreading velocity is significantly lower than the spreading velocity on the uncovered heater from the very beginning. At $t = 14$ ms, the velocity decreases abruptly and the second spreading phase begins, but is not completed after $250$ ms.  If scenario (I) occurs, then the spreading radius reaches a larger maximum compared with the case with the uncovered heater (see Fig.~\ref{BN-Rad}, top). This means that the liquid imbibes through the mat radially outward even after the kinetic energy has completely dissipated. In this scenario, a very high heat flux is transferred to the liquid owing to the contact between the liquid and the hot surface. The liquid drop evaporates quickly, and marginal shrinkage of the spreading radius owing to the transition of the liquid to the vapor phase is observed. It should be noted that the procedure of determination of the spreading radius relies solely on the heat flux data, it is impossible to distinguish between the liquid spead over the coating and the liquid imbibed into the coating.}

The maximum spreading radius for scenario (II) is smaller than that for the uncovered heater (see Fig.~\ref{BN-Rad}, bottom). The high spreading velocity phase lasts approximately $20$ ms. During the low spreading velocity phase, the liquid is imbibed into the pores of the nanofiber mat, leading to a decrease in the temperature difference between the heater and liquid over time. Therefore, the imbibition of the liquid through the nanofiber mat leads to the gradual spreading of the liquid over the surface.

\begin{figure}
\centering\includegraphics[width=0.6\linewidth]{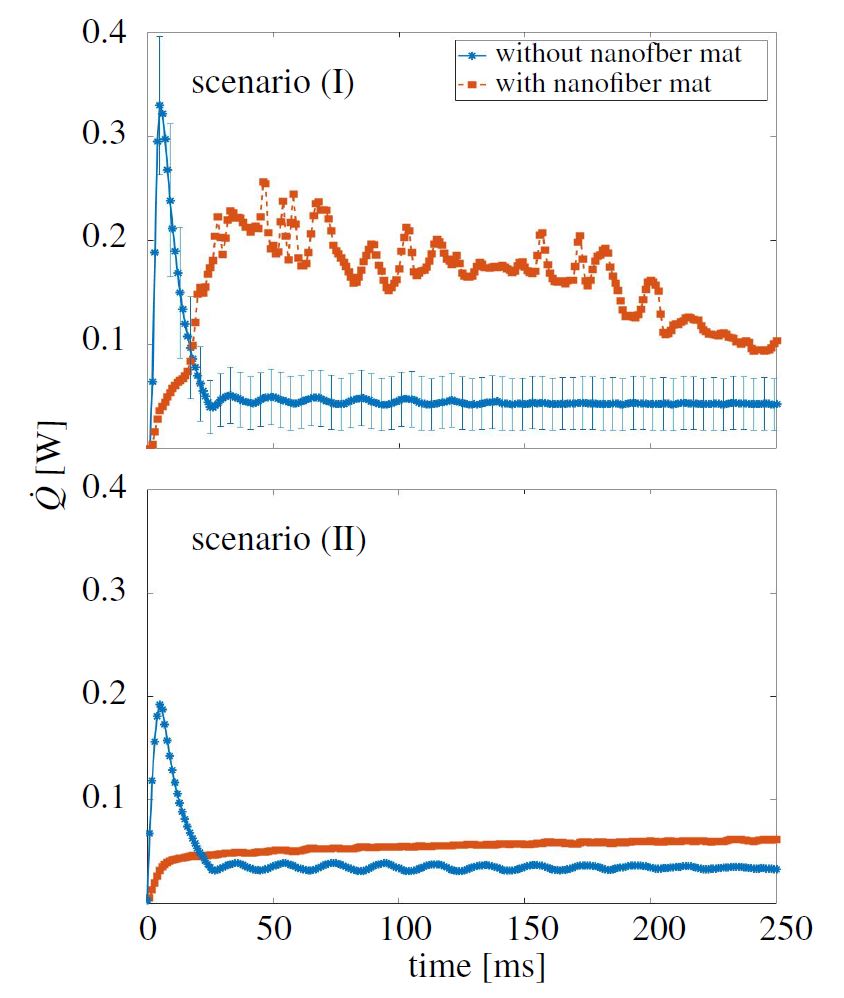}
\caption{Temporal evolution of the heat flow on a heater without and with the nanofiber mat for scenarios (I) and (II) for the parameters listed in Table \ref{Sen}.}
\label{BN-HF}
\end{figure}

Figure \ref{BN-HF} shows the temporal evolution of the heat flow for scenarios (I) and (II) for a heater without and with the nanofiber mat. In the case of the uncovered heater, a huge amount of heat is transferred to the liquid drop at the initial stage of impact when the cold liquid drop makes direct contact with the heater surface. If the heater is covered with the nanofiber mat and scenario (I) occurs, then a high heat flux and a corresponding heat flow comparable to but lower than the maximum heat flow transferred from the uncovered heater for a much longer time period can be observed (see Fig.~\ref{BN-HF}, top). This scenario is characterized by strong fluctuations of the heat flow, since the contact between the liquid and the heater substrate changes dynamically.

If scenario (II) occurs, then the heat flow follows the same trend as the spreading radius, and no maximum heat flow is identified. In this scenario, the heat flow is much lower (almost a quarter) than the maximum heat flow measured in the case of the uncovered heater (see Fig.~\ref{BN-HF}, bottom). Starting at $t = 22$ ms, the heat flow in scenario (II) exceeds the heat flow on the uncovered heater (during the sessile drop evaporation phase).


\begin{figure}
\centering\includegraphics[width=0.6\linewidth]{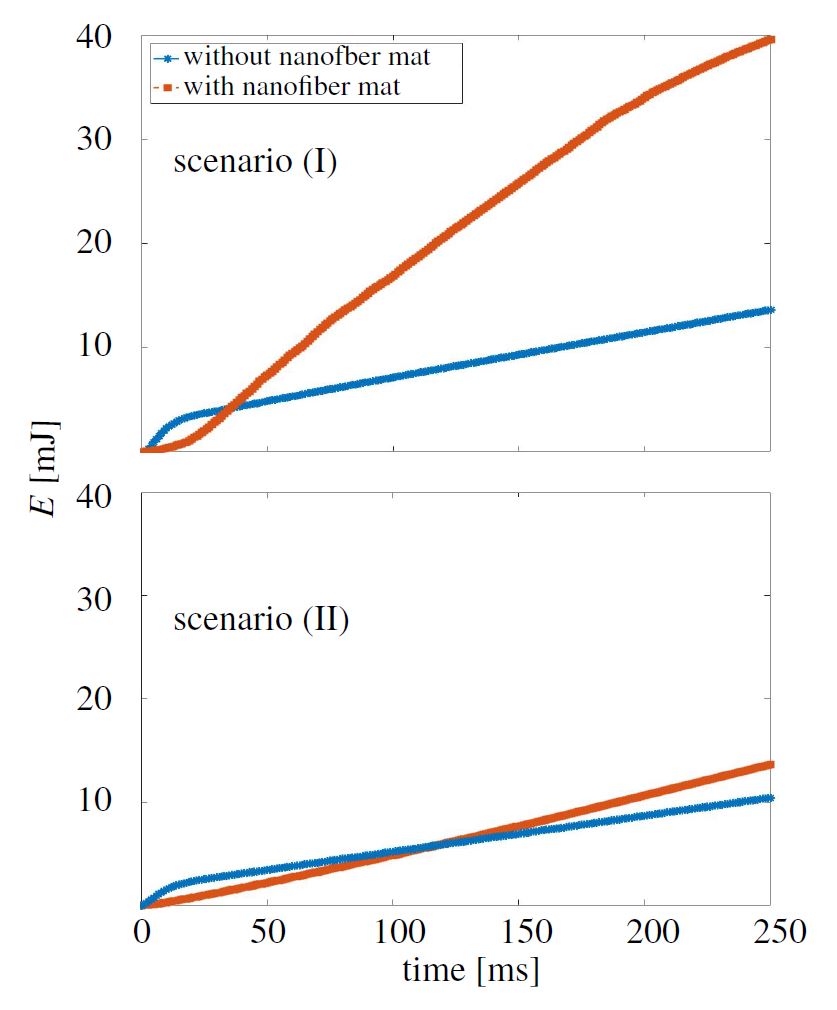}
\caption{Temporal evolution of the cumulative heat on a heater without and with the nanofiber mat for scenarios (I) and (II) for the parameters listed in Table \ref{Sen}.}
\label{BN-H}
\end{figure}

 

The temporal evolution of the cumulative heat in scenarios (I) and (II) for a heater without and with the nanofiber mat is presented in Fig.~\ref{BN-H}. The drop impact on the uncovered heater is accompanied by a sharp slope of the cumulative heat at the early stage of impact, owing to the high heat flux in this time period. Afterward, the cumulative heat increased almost linearly, as the transferred heat flow remained roughly constant during the sessile drop evaporation phase. 
 
In the case where the heater is covered with the nanofiber mat and scenario (I) is realzed, the cumulative heat rapidly rises apart from the initial phase (until $t = 17$ ms), which is attributed to the high heat flux transferred through the large drop footprint at each time step. If scenario (II) occurs, then the cumulative heat still rises, but not as steep as in scenario (I). Regardless of whether scenario (I) or (II) occurs, the cumulative heat or the energy transferred from the covered heater exceeds that of the uncovered heater after a specific time ($t = 30$ ms for scenario (I) and $t = 125$ ms for scenario (II)). This is associated with a larger drop footprint owing to the presence of the nanofiber mat and suppression of the drop receding phase. It can be generally stated that the presence of the nanofiber mat enhances the energy transferred to the liquid, despite the lower local heat flux at the initial stage of impact. 

\subsection{Influence of wall superheat}

\begin{figure}
\centering\includegraphics[width=0.6\linewidth]{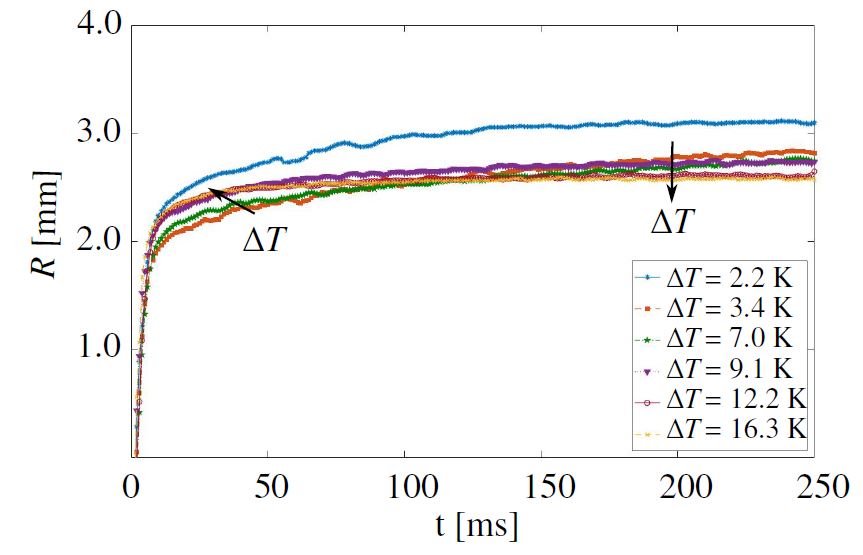}
\caption{Temporal evolution of the spreading radius for various wall superheats ($D_0 = 0.95$~mm, $u_0 = 0.45$~m~s$^{-1}$, and $h = 22$~{\textmu}m).}
\label{DT-Rad}
\end{figure}

\begin{figure}
\centering\includegraphics[width=0.6\linewidth]{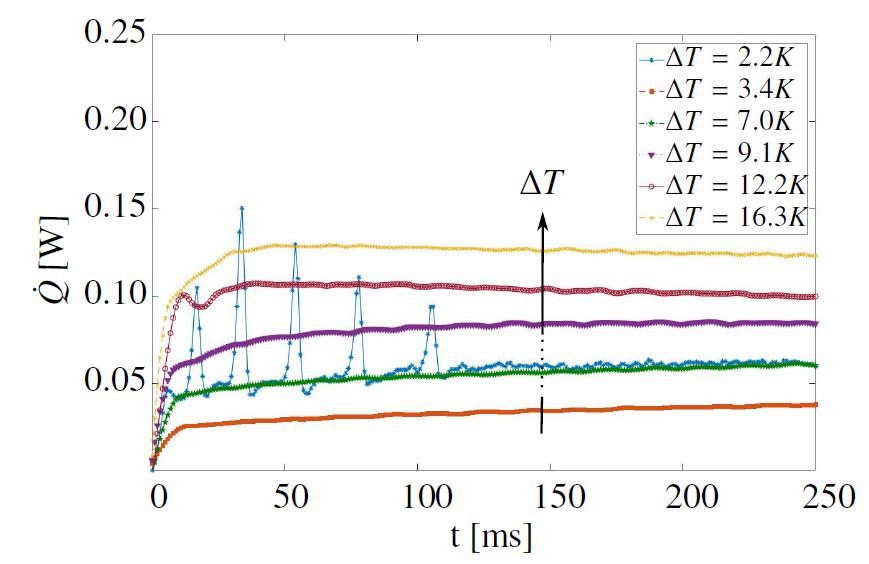}
\caption{Temporal evolution of the heat flow for various wall superheats ($D_0= 0.95$~mm, $u_0= 0.45$~m~s$^{-1}$, and $h = 22$~{\textmu}m).}
\label{DT-HF}
\end{figure}

\begin{figure}
\centering\includegraphics[width=0.6\linewidth]{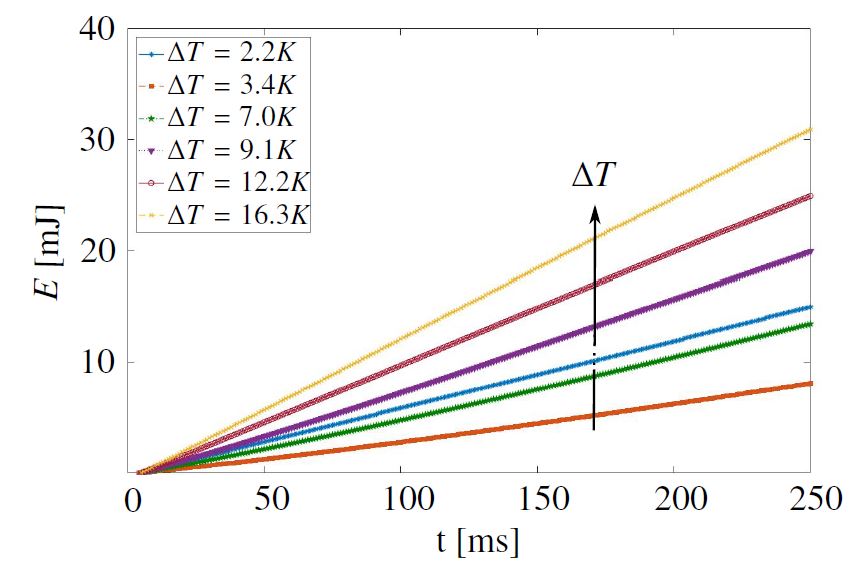}
\caption{Temporal evolution of the cumulative heat flow for various wall superheats ($D_0= 0.95$~mm, $u_0= 0.45$~m~s$^{-1}$, and $ h = 22$~{\textmu}m)}
\label{DT-H}
\end{figure}

\vspace*{0.3cm} 
Figures \ref{DT-Rad}, \ref{DT-HF}, and \ref{DT-H} show the temporal evolution of the spreading radius, heat flow, and cumulative heat during the drop impact on a surface covered with the nanofiber mat at wall superheats ranging from $2.2$ to $16.3$ K. Scenario (II) occurs in all wall superheats except at $\Delta T= 2.2$ K. It can be observed that at $\Delta T= 2.2$ K (scenario (I)), the spreading radius after the first 10 ms of spreading is higher than that for all cases, for which scenario (II) is realized. The heat flow for the case  $\Delta T= 2.2$ K shows an oscillating bahavior with several sharp peaks (Fig.~\ref{DT-HF}). Consequently, higher heat flow and cumulative heat are realized at this temperature compared with those at $\Delta T= 3.4$ and $7.0$ K (Fig.~\ref{DT-H}). The particular behavior of the spreading radius, heat flow, and cumulative heat at $\Delta T = 2.2$ K can be explained by the transition from scenario (I) to (II).

\begin{figure}
\centering\includegraphics[width=0.6\linewidth]{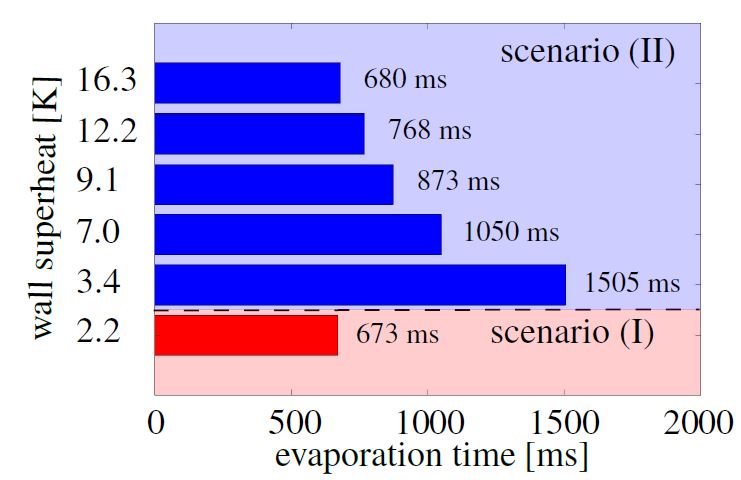}
\caption{Duration for entire drop evaporation for different values of wall superheat)}
\label{Time}
\end{figure}



The time required for complete drop evaporation according to the wall superheat is presented in Fig.~\ref{Time}. In the case $\Delta T= 2.2$ K for which scenario (I) occurs, the drop completely evaporated after $t =673$ ms, which is slightly quicker than the evaporation time at $\Delta T= 16.3$ K ($t =680$ ms). If the wall superheat increases to $3.4$ K, then scenario (II) is realized, and the duration of complete drop evaporation increases to $1505$ ms. The transition from scenario (I) to (II) is accompanied not only by a smaller spreading radius, lower heat flow, and cumulative heat, but also by a drastically longer duration of drop evaporation. If the wall superheat increases further, then the evaporation time for the entire drop decreases gradually. 

\begin{figure}
\centering\includegraphics[width=1\linewidth]{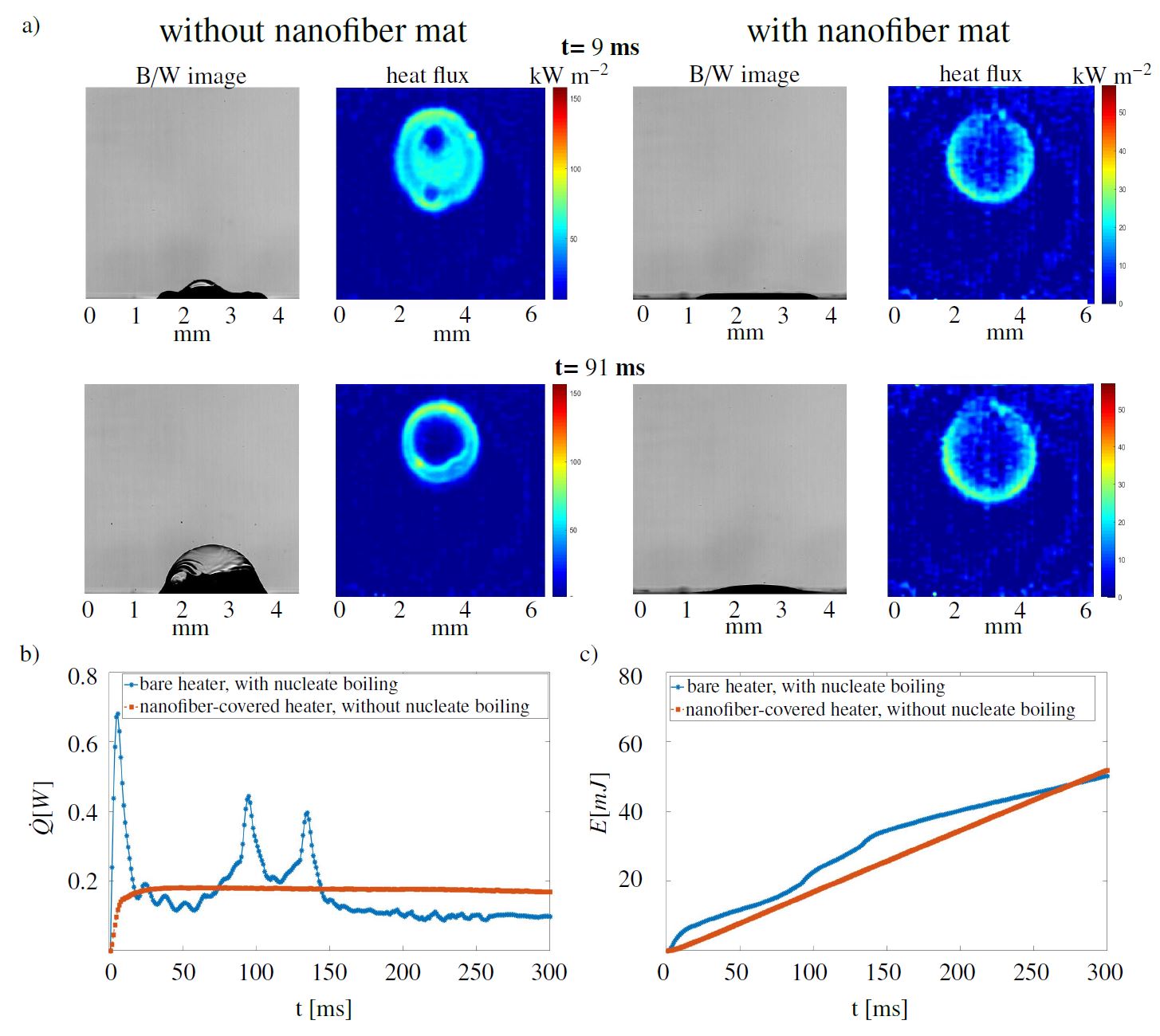}
\caption{a) B/W images and heat flux fields at $t= 9$ and $91$ ms for $\Delta T= 26.6$ K ($D_0= 0.95$~mm and $u_0= 0.45$~m~s$^{-1}$). Temporal evolution of the b) heat flow and c) cumulative heat during drop impact on an uncovered heater with nucleate boiling and a nanofiber-coated heater without nucleate boiling}
\label{NuBo}
\end{figure}
 
Figure \ref{NuBo} shows the B/W and IR images during drop impact on a surface with a high wall superheat ($\Delta T= 26.6$ K) for two similar times. Nucleate boiling can be clearly observed in the case of the uncovered heater. However, the nucleate boiling is suppressed if the heater is covered with the nanofiber mat. It can be concluded that the nanofiber-coated surface shifts the onset of nucleate boiling toward larger wall superheats. 

In Figure \ref{NuBo}(b) and (c), the heat flow and cumulative heat during drop impact on a bare heater (with nucleate boiling) and a nanofiber-coated heater (without nucleate boiling) are compared to each other. The second and third maxima in Figure \ref{NuBo}(b) during nucleate boiling arise from the bubble nucleation and growth. It can be observed that despite the high heat flow and the corresponding sharp increase in cumulative heat during nucleate boiling at the early stage of the impact, the cumulative heat for the nanofiber-coated heater exceeded that of the uncovered heater at $t= 277$ ms owing to the large drop footprint. Therefore, more energy is removed from the noanofiber-covered surface compared with the uncovered surface on which nucleate boiling occurs.


\subsection{Influence of impact velocity}

The experiments have been performed at the impact velocities of $0.45$, $0.54$, and $0.58$~m~s$^{-1}$. It is instructive to present the results in this section in a dimensionless form, altough the increase of the impact velocity at a constant drop diameter leads to increasingof both Reynolds and Weber numbers. We introduce the dimensionless time $\tau$ using the inertial time scale $D_0/u_0$. The spreading ratio is defined as follows:

\begin{equation}\label{eq:S}
S= \frac {2R} {D_0}.
\end{equation}

The dimensionless heat flow is defined as

\begin{equation}\label{eq:Q-dim}
Q^*= \frac {6\dot{Q}} {\pi \rho D_0^2u_0h_{lv}},
\end{equation}

and the dimensionless cumulative heat is defined as the ratio of the cumulative heat and the heat needed to completely evaporate the drop:

\begin{equation}\label{eq:E-dim}
E^*= \frac {6E} {\pi \rho D_0^3h_{lv}}.
\end{equation}

\begin{figure}
\centering\includegraphics[width=0.6\linewidth]{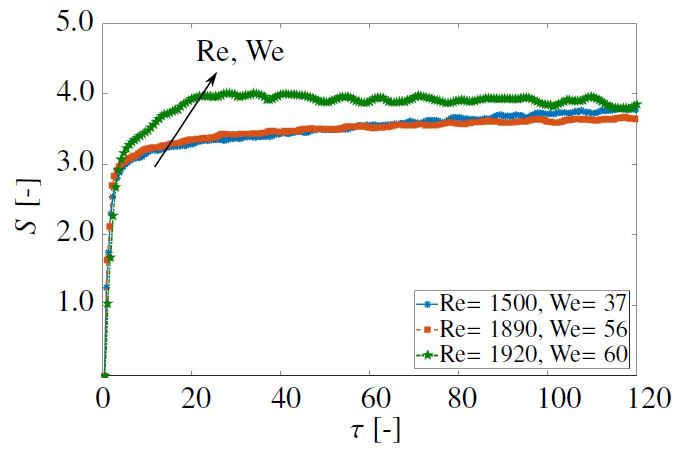}
\caption{Spreading ratio versus dimensionless time for varior Re and We (Ja=0.12).}
\label{V-S}
\end{figure}

\begin{figure}
\centering\includegraphics[width=0.6\linewidth]{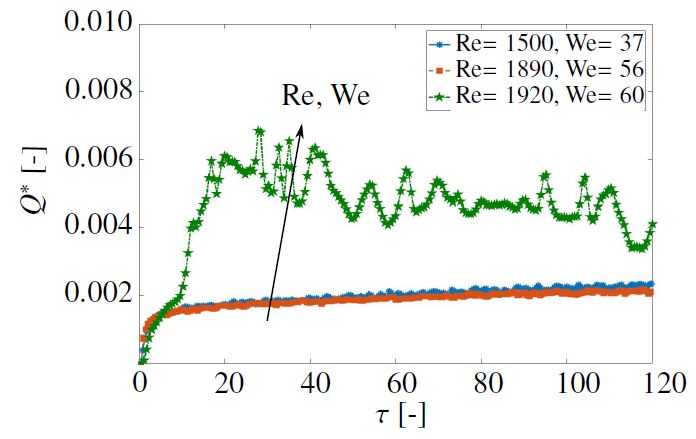}
\caption{Dimensionless heat flow versus dimensionless time for varior Re and We (Ja=0.12).}
\label{V-HFnd}
\end{figure}

\begin{figure}
\centering\includegraphics[width=0.6\linewidth]{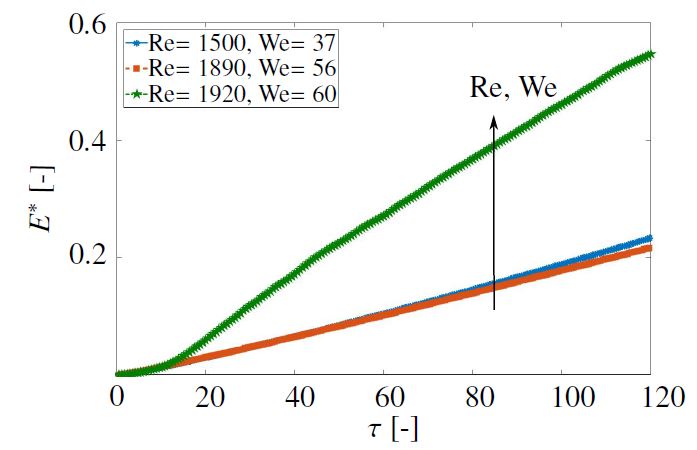}
\caption{Diensionless cumulative heat versus dimensionless time for varior Re and We (Ja=0.12).}
\label{V-Hnd}
\end{figure}

The temporal evolution of the spreading ratio, diensionless heat flow, and dimensionless cumulative heat at distinct impact velocities (and, therefore, disting values of Re and We) are shown in Figs. \ref{V-S}, \ref{V-HFnd}, and \ref{V-Hnd}, respectively. The initial evolution of the spreading ratio and the dimensionless heat transfer parameters are approximately similar for all three velocities, which agrees well with the results published earlier \cite{Gholijani1}. Moreover, for the two lower values of the impact velocities, the evolution of the spreading ratio, of dimensionless heat flow and, therefore, of the dimensionless cumulative heat, are nearly identical. 
The spreading ratio for the highest value of the velocity exceeds that measured at lower velocities. The dimensionless heat flow at the highest value of the velocity is highly fluctuating and exceeds significatly over the heat flow at lower impact velocities. This indicates that the drop impact with the velocity of $0.58$~m~s$^{-1}$ corresponds to scenario (I), and the two other cases correspond to scenario (II). If the impact velocity is sufficiently high, then the liquid drop expels the vapor trapped between the pores of the nanofiber mat radially outward, and establishes contact with the hot heater surface (scenario (I)). As described in section \ref{Nan}, this phenomenon slightly increases the spreading radius and leads to an increase in the heat flow and cumulative heat. 

\begin{figure}
\centering\includegraphics[width=0.8\linewidth]{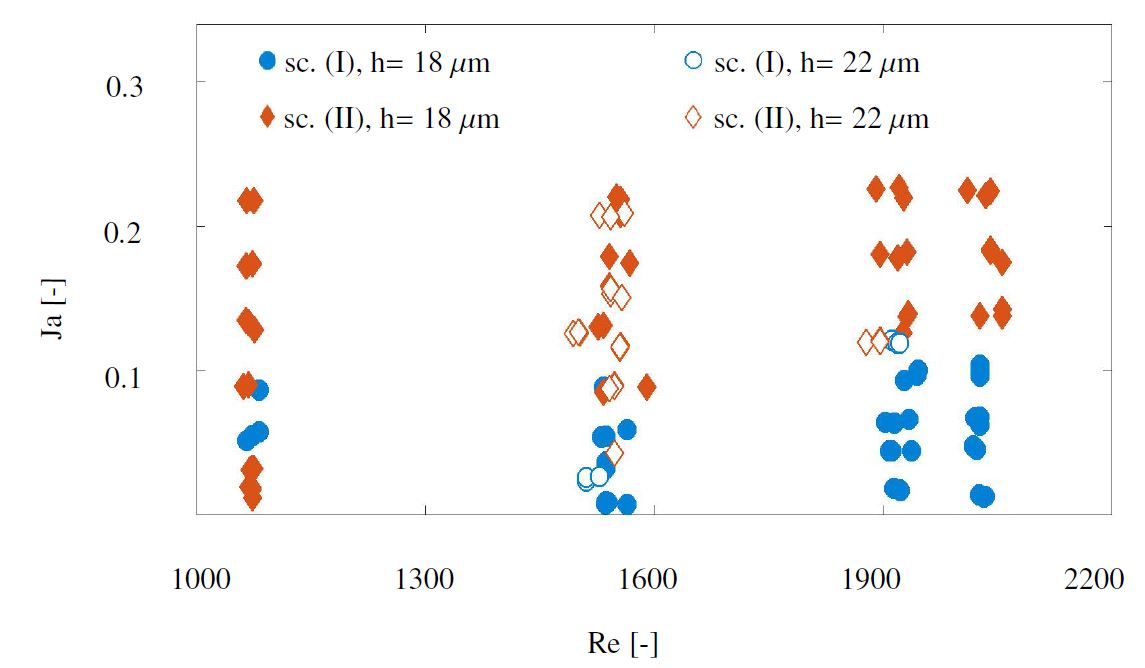}
\caption{Map of scenarios with dependence on Ja and Re numbers ($ h = 18 ~ $ and $ 22  $~{\textmu}m, $1.2$ K $\leqslant \Delta T \leqslant 17$ K, and $0.32$~m~s$^{-1} \leqslant u_0 \leqslant 0.58$~m~s$^{-1}$)}
\label{Map}
\end{figure}

Figure \ref{Map} depicts the regime map of scenarios that depend on the wall superheat and impact velocity, presented in dimensionless form using the Ja and Re numbers for the mat thicknesses of 18 and 22~{\textmu}m. Scenario (I) is mostly observed at low Ja and large Re numbers, at which the liquid drop penetrates the nanofiber mat. According to the report given in \cite{Sahu2} on the impact of liquid drops on different electrospun nanofiber membranes, a threshold impact velocity exists above which the liquid penetrates the membrane. This threshold is determined by competition between the inertia of the impacting drop and the viscous stresses within the mat. 

Within the experimental range, it can be observed that at high wall superheats or Ja numbers, scenario (II) occurs independent of the impact velocity or Re number. This is mainly attributed to the high evaporation rate at large wall superheats, which prevents the liquid drop from making contact with the heater surface. This scenario also occurs at the lowest Ja and Re numbers, at which the inertial force of the drop is not high enough to expel the vapor inside the pores radially outward. Then, the increasing of Ja leads to a regime change from scenario (II) to (I), and then back from (I) to (II). A similar observation has been made in \cite{Srikar}. It has been observed that, wenn a drop of a liquid is gently deposited onto a nanofiber mat, it can stay over a long time over a coating without a penetration. Heatingthe substrate can accelerate the penetration of the liquid into the mat.

It can be seen in Figure \ref{Map} that the regime maps for $h = 18$~{\textmu}m and $h =22$~{\textmu}m are qualitatively very similar. However, the transition from scenario (I) to scenario (II) takes place at a slightly lower value of Ja for a thicker mat, if the Reynolds number stays the same.

\subsection{Influence of mat thickness}

\begin{figure}
\centering\includegraphics[width=0.6\linewidth]{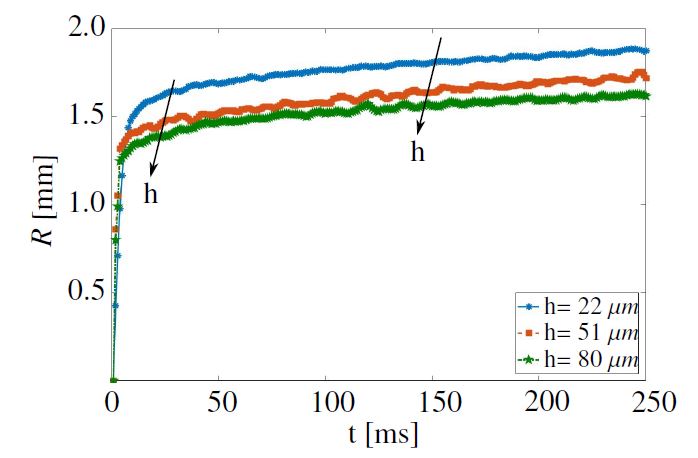}
\caption{Temporal evolution of the spreading radius for various mat thicknesses ($D_0= 0.95$~mm, $u_0= 0.45$~m~s$^{-1}$, and $\Delta T= 7.0$ K)}
\label{Thick-Rad}
\end{figure}

\begin{figure}
\centering\includegraphics[width=0.6\linewidth]{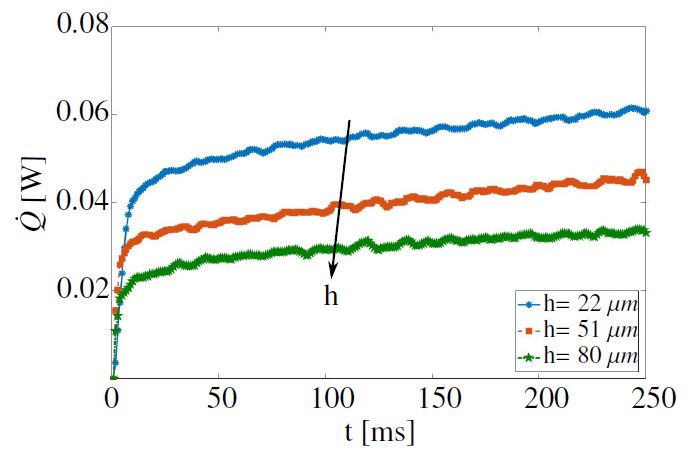}
\caption{Temporal evolution of the heat flow for various mat thicknesses ($D_0= 0.95$~mm, $u_0= 0.45$~m~s$^{-1}$, and $\Delta T= 7.0$ K)}
\label{Thick-HF}
\end{figure}

\begin{figure}
\centering\includegraphics[width=0.6\linewidth]{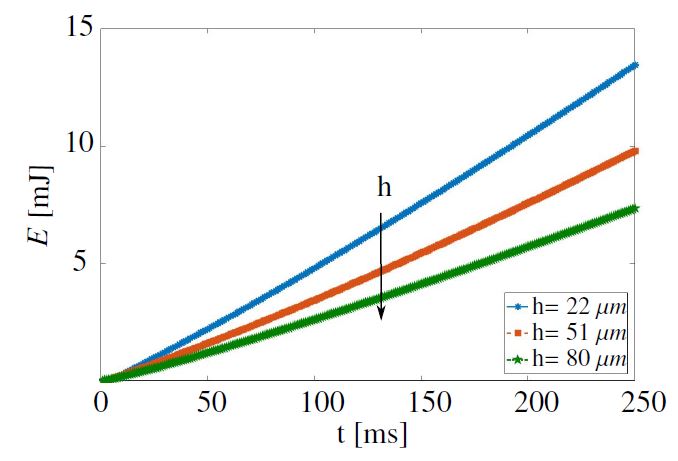}
\caption{Temporal evolution of the cumulative heat for various mat thicknesses ($D_0= 0.95$~mm, $u_0= 0.45$~m~s$^{-1}$, and $\Delta T= 7.0$ K)}
\label{Thick-H}
\end{figure}



\vspace*{0.3cm}
Figures \ref{Thick-Rad}, \ref{Thick-HF}, and \ref{Thick-H} show the temporal evolution of the spreading radius, heat flow, and cumulative heat for nanofiber mat thicknesses of $22$, $51$ and $80$~{\textmu}m ata fixed impact velocity of 0.45 ~m~s$^{-1}$ and a fixed superheat of 7.0 K. In the studied range of thicknesses and at the fixed impact and thermodynamic parameters, only scenario (II) occurs. Thinner nanofiber mats lead to a larger spreading radius, since the dissipation of kinetic energy is lower. If the mat is thinner, higher heat flow and cumulative heat are transferred to the drop. This is mainly attributed to the bigger spreading radius. 


\section{Conclusions}
In this study, the drop dynamics as well as local and overall heat transfer during drop impingement on a hot surface covered with an electrospun nanofiber mat were investigated. The influence of wall superheat ($2.2$--$16.3$ K), drop impact velocity ($0.45$--$0.58$~m~s$^{-1}$), and nanofiber mat thickness ($18$--$80$~{\textmu}m) on the hydrodynamics and heat transfer were investigated. The drop impingement on the bare surface is divided into three subsequent phases: (i) spreading phase, (ii) receding phase, and (iii) sessile drop evaporation phase. However, the drop receding phase was suppressed by the nanofiber mat. 

Two scenarios dependent on the wall superheat and impact velocity were identified. Scenario (I) occurred at low wall superheats and high impact velocities, in which the liquid drop completely penetrated the pores of the nanofiber and reached the solid heater surface. Scenario (II) occurred at large wall superheats, at which a large amount of vapor was trapped inside the nanofiber pores. This scenario occurred at small impact velocities in which the inertial force was not high enough to expel the generated vapor radially outward. The vapor inside the nanofiber pores prevented the liquid drop from reaching the solid surface, which led to a lower heat flux compared with scenario (I). The transition from scenario (I) to (II) leads to a smaller spreading radius, a lower heat flow and cumulative heat, and a longer time duration of drop evaporation.

The results also confirmed the shift in the onset of nucleate boiling toward the higher wall superheats if the surface is covered with the nanofiber mat. Thinner nanofiber mats lead to a larger spreading radius, heat flow, and cumulative heat transferred through the liquid-solid interface. 

In the future, the effects of mat thickness, wall superheat, and impact velocity on the transition from scenario (I) to (II) should be studied in details.

\section*{Acknowledgment}
The authors kindly acknowledge the financial support provided by the Deutsche Forschungsgemeinschaft (DFG, German Research Foundation) in the framework of CRC-TRR 75, Project-ID 84292822, subproject C02, and CRC 1194, Project-ID 265191195, subproject A04. The authors would also like to thank M. Heinz for the information exchange associated with the generation of the nanofiber mat.

\bibliographystyle{unsrtnat}  

\end{document}